# The concentration and health risk assessment of heavy metals and microorganisms in the groundwater of Lagos, Southwest Nigeria

**Tajudeen O. Yahaya[1,✉] , Esther O. Oladele[2], Itunuoluwa A. Fatodu[3], Abdulmalik Abdulazeez[1], Yusuf I. Yeldu[1]**

1. Department of Biology, Federal University Birnin Kebbi, P.M.B 1157, Birnin Kebbi, Nigeria
2. Biology Unit, Distance Learning Institute, University of Lagos, Nigeria
3. Department of Environmental Science and Toxicology, National Open University of Nigeria, Lagos

## ABATRACT

Contaminated water is increasingly linked to diseases worldwide, necessitating the safety evaluation of the sources of domestic and drinking water in every locality. The present study aimed to assess water safety in terms of lead (Pb), nickel (Ni), cadmium (Cd), copper (Cu), chromium (Cr), zinc (Zn), and microorganisms in the borehole and well water in Iwaya, Makoko, and Ilaje in Lagos, Nigeria. Water samples were subjected to atomic absorption spectroscopy (AAS) and microbiological examinations using standard protocols and compared with the World Health Organization (WHO) permissible limits. The average daily ingestion (ADI) and hazard quotient (HQ) of the selected heavy metals were also calculated using standard formulas. The AAS indicated that the borehole and well water of the three locations contained non-permissible levels of Pb, Ni, and Cd (only the well water in Ilaje), while Zn, Cu, and Cr were normal. In addition, the microbiological examinations showed that the borehole and well water of the three locations contained abnormal bacteria and coliform counts (well water only). The ADI and HQ of the selected heavy metals were less than one, which is the threshold at which a substance is considered safe for consumption. Water could pose some health risks, and the consumers in high-risk areas should consider water treatment before consumption.

**Keywords:** Average Daily Ingestion, Bacteria, Boreholes, Hazard Quotient, Lead

## Introduction

Water is the most important source of life on earth. The availability of safe drinking water is a basic human right, as well as an index of healthy living.[1] However, water is increasingly contaminated worldwide, accounting for over 1.8 million deaths yearly.[2] Children are most frequently affected by contaminated water, and a minimum of 525,000 children worldwide die every year due to diarrheal illnesses, most of which are caused by contaminated water and poor sanitation and personal hygiene.[1, 3] Some of the main causes of high water contamination are urbanization, industrialization, population growth, poor technologies, wars, and climate change.[1]

Heavy metals are among notable water contaminants, which are generally toxic metals with a specific density ($>5$ g/cm$^3$).[4] Heavy metals of public health concern regarding water contamination include arsenic (As), cadmium (Cd), nickel (Ni), mercury (Hg), chromium (Cr), zin c (Zn), copper (Cu), and lead (Pb).[5] The main sources of these metals in water are soil erosion, weathering, mining, industrial wastewaters, urban runoff, sewage discharge, municipal wastes, and agrochemicals.[6] Heavy metals generate reactive oxygen species in living organisms, thereby causing oxidative damage.[5]

Microorganisms are another group of water contaminants of public health concern. Overall, 500 waterborne microorganisms have been identified, which are classified as viral, bacterial, parasitic protozoan, and fungal pathogens.[7] The most common sources of exposure to microbes in water are human and animal excreta.[8] Microorganisms invade the cells and compromise the immune system or release toxins.[9] Notably, some of the microorganisms and heavy metals mentioned earlier (particularly Cu, Ni, Cr, and Zn) are found naturally in minute quantities in biological systems and are essential to the normal function of the body,[10, 11] and they become harmful only at high doses or due to environmental factors,[11] such as contaminated water. Therefore, the constant monitoring of the levels of these heavy metals and microorganisms in various water sources in every locality is critical for disease prevention.

Groundwater is a major source of water supply, which could be found in form of borehole and well water. At the minimum, groundwater accounts for 50% of the water supplies worldwide.[12] In Lagos, which is located in the southwest of Nigeria, groundwater is the primary source of water for drinking and domestic purposes[13] due to the inadequacy of pipe-borne water in some regions of the state, similar to many other developing countries. Considering the cosmopolitan nature of this state, the periodic monitoring of groundwater is essential to the prevention of diseases and disease outbreaks through the groundwater supply.

To the best of our knowledge, a monitoring study with the mentioned objectives has not been carried out in Iwaya, Makoko, and Ilaje, which are among the densely populated areas of Lagos. The few studies conducted in some localities close to these three areas have only evaluated some selected heavy metals without determining the health hazards and risks of the daily consumption of the contaminated water. Additionally, most of the studies in this regard have not considered the microbial contents of water, and none have combined heavy metal and microbial assessment. The present study aimed to determine the levels and safety status of selected heavy metals and microorganisms in the borehole and well water in Iwaya, Makoko, and Ilaje. The findings could help determine the suitability of borehole and well water in these areas for drinking.

**Materials and Methods**
*Description of the study areas*

This study was carried out in Ilaje, Iwaya, and Makoko areas in Lagos, located in the southwest of Nigeria (Fig. 1). Lagos is the capital of Lagos State with the latitudes of 6°37′N and 6°70′N and longitudes of 2°70′E and 4°35′E.[14] It covers an area of approximately 3,577 square kilometers, of which land constitutes 2,798 square kilometers and water covers 779 square kilometers.[14] Lagos is bordered by the Republic of Benin on the west, Ogun State on the east and north, and the Atlantic Ocean on the south.[14] Lagos is among the most populated and rapidly-growing cities in the world.[15] The city is the economic hub of Nigeria and most industrialized area in the country, responsible for large-scaled waste generations in the city. The state is characterized by tropical vegetation with many water bodies (e.g., lagoons, rivers, and creeks), a mainly humid climate, and a short dry season.

Among the notable places in Lagos are Ilaje in Shomolu local government and Makoko and Iwaya in Yaba local government. Shomolu is in the north of Lagos and characterized by overcrowding, poor housing, and inadequate sanitation. Most of the residents are artisans and particularly known for printing works. Yaba is in Lagos Mainland and has several educational institutions, the most notable of which are the Queen's College, the Nigerian Institute of Medical Research, the Yaba

College of Technology, and the University of Lagos. Apart from the presence of heavy artisan works and small-scaled industries, Yaba is home to Tejuosho Market which is one of the prominent markets in Lagos. Overall, numerous an thropogenic activities take place in these areas with potential environmental effects, particularly on drinking and domestic water, which require constant monitoring.

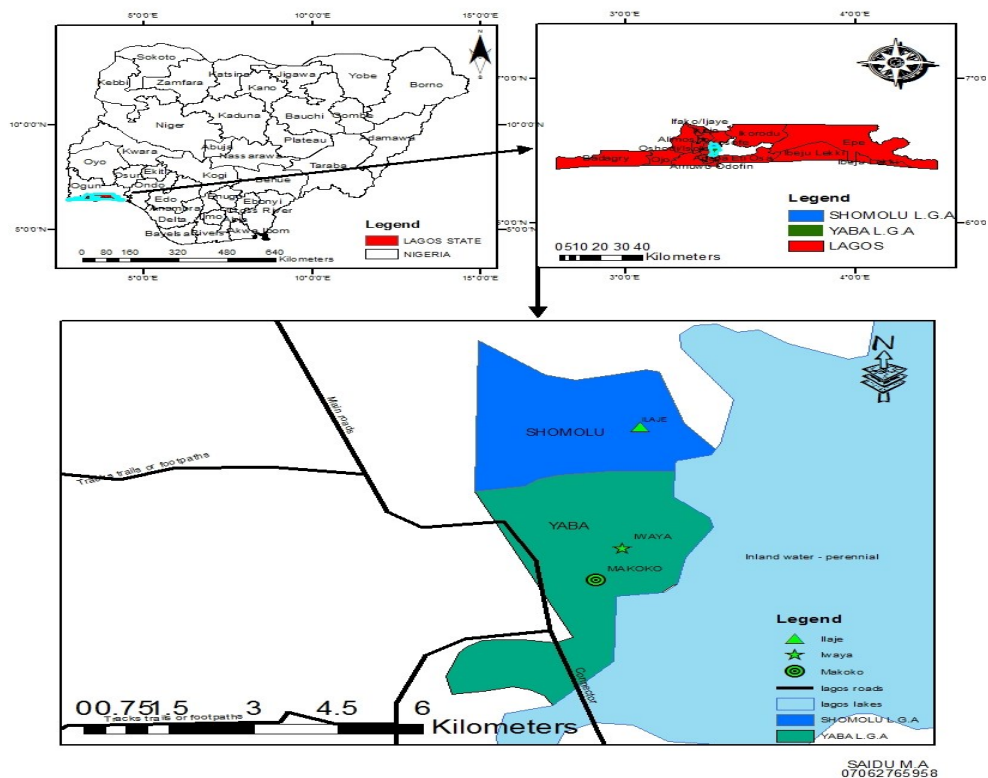

Fig. 1. Location map of study areas (ArcGIS 10.3 software)

### Water sample collection and preparation

In total, 20 samples of borehole water and 20 samples of well water were collected randomly from Ilaje, Maroko, and Iwaya during February-June 2020 and placed in clean, pre-sterilized plastic containers. The samples were covered tightly and moved to the laboratory where they were stored in desiccators before heavy metal and microbiological analyses.

### Heavy metal analysis

The water samples were digested and analyzed for heavy metals using the method proposed by Yahaya et al.[16] One milliliter of each of the test substances was transferred to a pre-washed 100-milliliter beaker containing analytical grade, 25 milliliters of aqua regia mixture (70% $HNO_3$ and HCl ratio: 3:1), and five milliliters of 30% $H_2O_2$. The mixture was digested in a digestion vessel at the temperature of 80 °C until obtaining a homogenous solution. Afterward, the solution was cooled, filtered through a Whatman No. 42 filter paper into a 50-milliliter volumetric flask, and diluted to the mark with deionized water. The filtrate was subjected to atomic absorption spectroscopy using a spectrophotometer (UNICAM, model: 969) to determine the concentrations of Cu, Pb, Cd, Cr, Ni, and Zn.

### Microbial analysis

The total bacterial counts were estimated using the membrane filtration technique as described by Brock.[17] To this end, 100 mL of each water sample was filtered through a sterile cellulose filter, and the filter was placed on a nutrient agar plate and incubated for 24 h at the temperature of 35 °C. The total bacterial colonies formed on the plate were estimated using a colony counter, and the membrane filtration technique was also used to estimate the coliform count. In addition, the two-step enrichment method was used for microbial growth.[18] Following that, the filters containing the bacteria were placed on an absorbent pad saturated with lauryl tryptose broth, incubated at the temperature of 35 °C for 2 h, transferred to an absorbent pad saturated with M-Endo media, and incubated for 22 h at the temperature of 35 °C. Sheen colonies were also observed and estimated by a colony counter.

### Quality control

The background contamination of the samples was evaluated to ensure the accuracy of the data. The blank samples were analyzed after five samples, and the analyses were performed in triplicate. The precision and accuracy of the analyzed heavy metals were assessed against standard reference materials for each element.

### Health risk assessment of the heavy metals

The health risk of the selected heavy metals in the borehole and well water samples was calculated based on the average daily ingestion (ADI) (Eq. (1)) and hazard quotient (HQ) (Eq. (2)) of the heavy metals.[19]

$$ADI = \frac{Cx \times Ir \times Ef \times Ed}{Bwt \times At} \qquad (1)$$

In Eq. (1), ADI shows the average daily ingestion per kilogram of the body weight, Cx is the concentration of the heavy metals in water, Ir represents the ingestion rate per unit time, Ef is the exposure frequency, Ed shows the exposure duration (equal to the life expectancy of a Nigerian resident), Bwt denotes the body weight, and At is the average time (Ed x Ef). The standard values and units for these parameters are presented in Table 1.[20]

$$HQ = \frac{ADI}{RFD} \qquad (2)$$

In Eq. (2), HQ represents the hazard quotient, and RFD shows the oral reference dose (mg/L/day) of the selected heavy metals (Table 2).[20] The HQ of less than one was considered non-toxic.[21]

Table 1. Standard values for calculating average daily ingestion of heavy metals

| Exposure factors | Units | Values |
|---|---|---|
| Concentration of metals in water (Cw) | mg/L | - |
| Exposure frequency (Ef) | Days/year | 365 |
| Ingestion rate (Ir) | L/day | 2 |
| Exposure duration (Ed) | Years | 55 |
| Average body weight (Bwt) | Kg | 65 |

Table 2. Oral reference doses (RFD) for Pb, Ni, Cd, Cr, Zn, and Cu in water

| Heavy metal | Values |
|---|---|
| Pb | 14 |
| Ni | 20 |
| Cd | 0.5 |
| Cr | 3,0 |
| Zn | 300 |
| Cu | 40 |

### Statistical analysis

Descriptive statistics were used to summarize the collected data from the sampling sites, which were expressed as mean and standard deviation (SD). The graphs were drawn using the Minitab software version 7.0.

## Results and Discussion
### Levels of the selected heavy metals in the water samples

Tables 3 and 4 show the levels of Pb, Ni, Cd, Cr, Zn, and Cu in the samples of borehole and well water obtained from Iwaya, Makoko, and Ilaje. In the borehole samples, Pb and Ni concentrations were above the permissible limits of the World Health Organization (WHO) in all the locations, while Cd concentration was only abnormal in Iwaya and Makoko, and the concentrations of Cr, Zn, and Cu were normal in all the locations (Table 3). In the well water samples, the levels of Pb, Ni, and Cd were above the recommended limits in the three locations, while the Cr, Zn, and Cu concentrations were within the permissible limits (Table 4). These findings are consistent with the results obtained by Ehi-Eromosele and Okiei[22] and Chika and Prince[23] who detected abnormal quantities of Pb, Ni, and Cd in the groundwater of some parts of Lagos. Furthermore, the mentioned findings are in line with the studies by Adelekan et al.[24] and Afolabi et al.[12] which indicated the permissible levels of Cu, Zn, and Cr in the groundwater of Lagos Central and Oworonshoki (Nigeria). Therefore, it could be inferred that the various types of heavy metals in the drinking water sources across the city are uniform. However, these heavy metals were detected at higher levels in the present study compared to the aforementioned studies. Anthropogenic activities, population density, and sanitary conditions vary across the state, which might have influenced the uneven distribution of the heavy metals. Specifically, the residents of Iwaya, Ilaje, and Makoko areas of Lagos State have lower income and lower education levels compared with the residents of locations considered in the previous studies. In this regard, this could have affected their environmental awareness. In a study to compare the level of drinking water pollution between two areas in Lagos, the more populated area was reported to have higher levels of heavy metals.[12] In the same vein, Iwaya, Ilaja, and Makoko are heavily populated an d crowded, and this could contribute to the high level of water contamination in these areas.

The abnormal concentrations of Pb, Ni, and Cd in the present study suggested that both the borehole and well water could pose health risks to the consumers in the areas if not treated. The abnormal concentrations of Pb may also cause blood pressure, vitamin D and calcium metabolism imbalance, neurological disorders, and multi-organ damage.[13] Long-term exposure to Cd could cause kidney damage, lung cancer, hypertension, and bone diseases.[13] Ni exposure has been reported to cause cardiovascular and renal disorders, as well as lung and nasal cancer.[25] Although Cu, Zn, and Cr were within the permissible limits in the water samples of the current research, they may however, bio accumulate to toxic levels in the body and give rise to various health issues. Long-term exposure to hexavalent Cr may also induce cancer [26] and Cu exposure could lead to gastrointestinal disorders and liver damage.[27] Zinc exposure has been shown to promote apoptosis in the brain.[28] The noticeable sources of heavy metals in the studied locations of the present study were brimmed with the leaching of artisans' workshops, dumpsites, and sewage of homes and small-scaled industries.

Table 3. Levels of selected heavy metals in borehole water in Iwaya, Makoko, and Ilaja

| Location | Pb | Ni | Cd | Cr | Zn | Cu |
|---|---|---|---|---|---|---|
| Iwaya | 0.25± 0.006 | 0.11±0.015 | ND | 0.01±0.00 | 2.12±0.004 | 0.55±0.0012 |
| Makoko | 0.21±0.006 | 0.13±0.010 | ND | 0.02±0.006 | 2.30±0.015 | 0.48±0.020 |
| Ilaje | 0.27±0.006 | 0.20±0.006 | 0.02±0.006 | 0.02±0.00 | 2.31±0.001 | 0.57±0.010 |
| WHO Limit[29] | ≤0.01 | ≤ 0.02 | ≤ 0.02 | ≤0.05 | ≤5.00 | ≤5.00 |

Values expressed as mean ± SD and mg/L; ND: not detected; WHO: World Health Organization

Table 4. Levels of selected heavy metals in well water in Iwaya, Makoko, and Ilaja

| Location | Pb | Ni | Cd | Cr | Zn | Cu |
|---|---|---|---|---|---|---|
| Iwaya | 0.25± 0.006 | 0.13±0.015 | 0.02±0.006 | 0.02±0.006 | 2.53±0.017 | 0.55±0.0025 |
| Makoko | 0.31±0.015 | 0.21±0.006 | 0.02±0.006 | 0.02±0.006 | 2.54±0.006 | 0.48±0.015 |
| Ilaje | 0.37±0.006 | 0.17±0.006 | ND | 0.03±0.010 | 2.50±0.015 | 0.57±0.006 |
| WHO Limit[29] | ≤0.01 | ≤ 0.02 | ≤ 0.03 | ≤0.05 | ≤5.00 | ≤1.00 |

Values expressed as mean ± SD and mg/L; ND: not detected; WHO: World Health Organization

## Health risk of the selected heavy metals in water

According to the information in Tables 5 and 6, the ADI of Pb, Ni, Cd, Cr, Zn, and Cu by the residents of Iwaya, Makoko, and Ilaje in the borehole and well water was within the respective recommended daily intake (RDI). As depicted in figures 2 and 3, the HQ of the heavy metals in the borehole and well water samples in the three locations were less than one, suggesting that the water may not cause severe health effects on the residents with an average life expectancy (55 years) in Nigeria. As for the residents that live beyond 55 years, the risk may become increasingly significant with age. This finding is consistent with the study by Adeniyi et al.[30] which indicated that the HQ of selected heavy metals in well and borehole water in an area in Lagos was less than one. In addition, Ukah et al.[31] reported no health risks in terms of selected heavy metal concentrations in approximately 76% of well and borehole water evaluated in some regions of Lagos. However, the results were inconsistent with the findings of Sogbanmu et al.[32] who reported the HQ of more than one for some heavy metals in borehole and well water in Ibeju-Lekki in Lagos.

Table 5. Average daily ingestion of heavy metals per person in borehole water in Iwaya, Makoko, and Ilaje

| Location | Pb | Ni | Cd | Cr | Zn | Cu |
|---|---|---|---|---|---|---|
| Iwaya | 0.008 | 0.0034 | ND | 0.0003 | 0.065 | 0.009 |
| Makoko | 0.006 | 0.006 | ND | 0.0006 | 0.071 | 0.007 |
| Ilaje | 0.008 | 0.004 | 0.0006 | 0.0006 | 0.071 | 0.011 |
| RDI[33] | 0.21 | 0.500 | 0.06 | 0.20 | - | 0.900 |

Values expressed as mg/day; ND: not detected; RDI: recommended daily intake

Table 6. Average daily ingestion of heavy metals per person in well water in Iwaya, Makoko, and Ilaje

| Location | Pb | Ni | Cd | Cr | Zn | Cu |
|---|---|---|---|---|---|---|
| Iwaya | 0.0076 | 0.004 | 0.00062 | 0.00062 | 0.078 | 0.017 |
| Makoko | 0.0095 | 0.0065 | 0.0012 | 0.0006 | 0.078 | 0.015 |
| Ilaje | 0.011 | 0.005 | ND | 0.0009 | 0.077 | 0.018 |
| RDI[33] | 0.21 | 0.500 | 0.06 | 0.20 | | 0.900 |

Values expressed as mg/day; ND: not detected; RDI: recommended daily intake

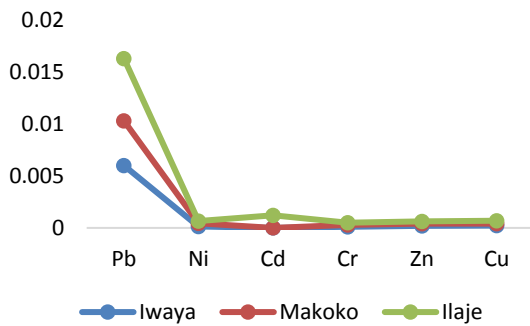

Fig. 2. Hazard quotient of heavy metals via ingestion of borehole water

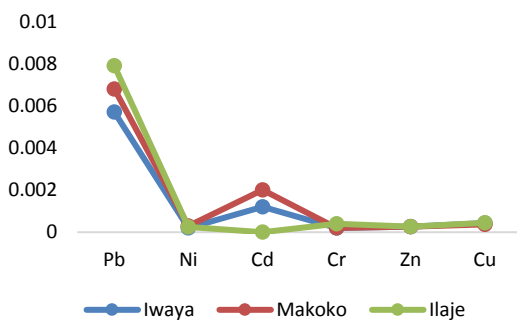

Fig. 3. Hazard quotient of heavy metals via ingestion of well water

As stated earlier, the observed differences in the mentioned studies with the present study may be due to the variations in the nature of anthropogenic activities, sanitary conditions, and population density of various areas. As reported by Sogbanmu et al.[32] the HQ of some heavy metals in the drinking water of Ibeju-Lekki is more than one; as Ibeju-Lekki is a metropolitan in Lagos, and this finding is considered to be significant.

### Microbial content of the water

Tables 7 and 8 show the bacterial and coliform counts in the borehole and well water samples collected from Iwaya, Makoko, and Ilaje. Accordingly, the borehole water samples of the three locations had abnormal bacterial counts, while coliforms were not detected (Table 7). Iwaya and Ilaje jointly had the highest bacterial count (12,000 CFU/mL), while Makoko had the lowest bacterial count (140 CFU/mL). In the well water samples of the three locations, bacterial and coliform counts were above the permissible limits (Table 8). Iwaya had the highest bacterial count (20,000 CFU/mL), followed by Makoko (18,000 CFU/mL) and Ilaje (12,000 CFU/mL). The coliform counts were highest in the well water samples collected from Ilaje (12 CFU/mL), followed by Makoko (11 CFU/mL) and Iwaya (10 CFU/mL). Therefore, it was concluded that the borehole and well water of the studied areas may not be suitable for drinking unless treated.

Several bacterial species produce nutrients, digest food, and boost the immune function.[34] On the other hand, some waterborne bacteria may cause diseases such as cholera, diarrhea, typhoid fever, and dysentery.[2] The presence of coliforms in the well water samples of the three locations indicated that the water was contaminated by environmental pollutants,[24] particularly fecal matters. Most coliforms are harmless, while certain strains of *Escherichia coli* 0157:H7, which are the most common fecal coliforms often found in animal feces and may cause diseases, especially diarrhea.[35] In previous studies, Egwari and Aboaba[36] and Adenekan and Ogunde[24] have also detected *E. coli* and enteric bacteria in several wells, boreholes, and lagoons in Lagos. However, total bacteria were detected at significantly higher counts in the present study compared to the other findings in this regard. Apart from sanitary conditions and anthropogenic activities that may vary across the state, seasonal variations might be an important factor involved in this issue. The current research was carried out in the wet season, and Egwari and Aboaba[36] have claimed that such a climate may favor the growth and sinking of microorganisms into the groundwater. The samples of the well water in the present study were generally more contaminated compared to the boreholes possibly because boreholes are deeper than wells. In the study conducted by Egwari and Aboaba,[36] shallow wells were reported to be more contaminated than deep wells. Notably, the probable sources of microbial contaminants in the water of the three locations in our study are the wastewaters discharged from homes, small-scaled industries, municipal wastes, and indiscriminate siting of wells and boreholes.

Table 7. Bacterial and coliform counts of borehole water in Iwaya, Makoko, and Ilaje

| Location | Bacteria count | Coliform count |
|---|---|---|
| Iwaya | 12000±1000 | ND |
| Makoko | 140.00±10 | ND |
| Ilaje | 12000±700 | ND |
| WHO Limit[37] | ≤100 CFU/mL | 0 CFU/100 mL |

Values expressed as mean ± SD and CFU/mL; ND: not detected; WHO: World Health Organization

Table 8. Bacterial and coliform counts of well water in Iwaya, Makoko, and Ilaje

| Location | Bacteria count | Coliform count |
|---|---|---|
| Iwaya | 20000±8000 | 10.00±2.00 |
| Makoko | 18000±9000 | 11.00±1.00 |
| Ilaje | 12000±580 | 12.00±2.00 |
| WHO Limit[37] | ≤100 CFU/mL | 0 CFU/100 mL |

Values expressed as mean ± SD and CFU/mL; ND: not detected; WHO: World Health Organization

### Recommendations

Based on the findings of the present study, the following suggestions are proposed:

- Residents are advised to practice personal hygiene.
- Residents should consider water treatment before consumption.
- Indiscriminate siting of boreholes and wells should be avoided.
- There is an urgent need for efficient waste management and environmental sanitation in the areas.
- There is an urgent need for public enlightenment in the areas regarding the health consequences of drinking contaminated water.

### Limitations of the study

The present study had a few limitations:

- Due to financial constraint, we could not assess all heavy metals and other toxic chemicals, which could be found in high concentrations in the water.
- Due to financial constraint, we could not cover all waterborne pathogens.

### Conclusion

According to the results, the borehole and well water in Iwaya, Makoko, and Ilaje contained abnormal l evels of Pb, Ni, and Cd, as well as permissible levels of Cu, Zn, and Cr. The ADI and HQ of the selected heavy metals in the three locations were less than one, which is the threshold at which a substance may be considered safe for consumption. Therefore, the selected heavy metals may not pose a significant health threat to the residents that live within the life expectancy of Nigeria (55 years). However, abnormal bacterial and coliform counts (well water only) were detected in the water, which indicated that the water may require proper treatment before it is considered safe for consumption.

**Conflicts of interest**
None declared.
**Acknowledgements**
Not applicable.